\documentclass[12pt,preprint]{aastex}
\usepackage{graphicx}
\usepackage{epsfig}
\usepackage{amsmath}
\usepackage{lscape}
\usepackage{color}

\newcommand{\psr}{PSR~J2021+4026}
\begin{document}
\slugcomment{}

\shorttitle{X-ray pulsation of PSR J2021+4026}
\shortauthors{Lin \& Hui et al.}

\title{Discovery of X-ray Pulsation from the Geminga-like Pulsar PSR~J2021+4026}

\author{L. C. C. Lin\altaffilmark{1}, C. Y. Hui\altaffilmark{2}, C. P. Hu\altaffilmark{3}, J. H. K. Wu\altaffilmark{4}, R. H. H. Huang\altaffilmark{4}, L. Trepl\altaffilmark{5},  J. Takata\altaffilmark{6}, K. A. Seo\altaffilmark{2}, Y. Wang\altaffilmark{6}, Y. Chou\altaffilmark{3} and K. S. Cheng\altaffilmark{6}}

\email{cyhui@cnu.ac.kr}

\begin{abstract}

We report the discovery of X-ray periodicity of $\sim$265.3~ms from a deep {\it XMM-Newton} observation 
of the radio-quiet $\gamma$-ray pulsar, PSR~J2021+4026, located at the edge of the supernova remnant G78.2+2.1 ($\gamma$-Cygni). 
The detected frequency is consistent with the $\gamma$-ray pulsation determined by the observation of 
{\it Fermi} Gamma-ray Space Telescope at the same epoch.  
The X-ray pulse profile resembles the modulation of hot spot on the surface of the neutron star. 
The phase-averaged spectral analysis also suggests that the majority of the observed X-rays have a thermal origin.
This is the \emph{third} member in the class of radio-quiet pulsars with the significant pulsations detected from both X-rays and $\gamma$-ray regimes. 
\end{abstract}

\keywords{Gamma rays: general --- pulsars: general --- ISM: individual objects: G78.2+2.1 --- pulsars: individual 
(PSR J2021+4026) --- X-rays: general --- radiation mechanisms: thermal}

\altaffiltext{1}{General Education Center, China Medical University, Taichung 40402, Taiwan} 
\altaffiltext{2}{Department of Astronomy and Space Science, Chungnam National University, Daejeon, South Korea} 
\altaffiltext{3}{Graduate Institute of Astronomy, National Central University, Jhongli 32001, Taiwan}
\altaffiltext{4}{Institute of Astronomy, National Tsing-Hua University, Hsinchu 30013, Taiwan}  
\altaffiltext{5}{Astrophysikalisches Institut and Universit$\ddot{a}$ts-Sternwarte, Universit$\ddot{a}$t Jena, Schillerg$\ddot{a}\beta$chen 2-3, 07745 Jena, Germany} 
\altaffiltext{6}{Department of Physics, University of Hong Kong, Pokfulam Road, Hong Kong, PRC}

\section{Introduction}
Before the launch of {\it Fermi} Gamma-ray Space Telescope, there was only one radio-quiet $\gamma$-ray pulsar has been known, namely Geminga 
(i.e. PSR~B0633+17 Bertsch et al. 1992).
The high sensitivity of Large Area Telescope (LAT) on board {\it Fermi} enables an efficient search of $\gamma$-ray pulsars 
(Abdo et al. 2008, 2010a; Ackermann et al. 2011). 
Soon after the commence of the mission, (Abdo et al. 2009a) has reported the detections of 16 pulsars with 13 of them remain to be radio-quiet after the dedicated radio pulsation search (Camilo et al. 2009; Abdo et al. 2010b). 
The population of radio-quiet $\gamma$-ray pulsars has been expanded to 22 based on the extensive searches of 
Saz Parkinson et al. (2010). 
Thanks to the improved search techniques provided by Kerr (2011) \& Pletsch (2011), in total, 
there are 31 radio-quiet $\gamma$-ray pulsars are currently known (Pletsch et al. 2012a,b).

For this class of pulsars, the lack of knowledge in the phase relationship between the $\gamma-$ray light curve and that in radio leads to an ambiguity in investigating their emission region and physical processes (see the discussion in Trepl 
et al. 2010). 
Therefore, pulsation searches in the other frequencies are fundamentally important for a better understanding to the physics of this class of pulsars.
So far, there are only Geminga (Halpern \& Holt 1992) and PSR~J0007+7303 (Lin et al. 2010; Caraveo et al. 2010) have their X-ray pulsations been firmly detected. 
In this work, we reported our results of X-ray periodicity search of another member in this class --- \psr. 

\psr\ is one of the brightest $\gamma$-ray pulsars, which is associated with the supernova remnant (SNR) G78.2+2.1, detected by {\it Fermi} LAT 
shortly after its operation (Abdo et al. 2009). 
The possible association with SNR suggests it is at a distance of $\sim1.5$~kpc (Trepl et al. 2010 and references therein). 
Its proximity makes it as an ideal target for multi-wavelength investigations. Several efforts have been devoted in searching its radio counterpart, no radio pulsar associated with \psr\ has been detected so far (Becker et al. 2004; Trepl et al. 2010; 
Ray et al. 2011). 

The timing ephemeris of \psr\ has been determined by {\it Fermi} LAT at different epochs 
(Abdo et al. 2009a,2010a; Ackermann et al. 2011; Ray et al. 2011; Weisskopf 2011) 
and no obvious glitch was detected between the epoch of MJD 54680 to MJD 55600. 
Its observed period and period derivative ($P=265$~ms and $\dot{P}=5.48\times10^{-14}$~s~s$^{-1}$) imply a spin-down age of $\sim77$~kyr, surface dipolar magnetic field strength of $\sim4\times10^{12}$~G and a spin-down power of $\sim10^{35}$~erg~s$^{-1}$.
At a distance of 1.5~kpc, its $\gamma$-ray conversion efficiency is not dissimilar from that of Geminga (Trepl et al. 2010). 

In X-ray regime, a previously unidentified X-ray source, 2XMM~J202131.0+402645, has been identified as the promising counterpart of \psr\ (Trepl et al. 2010; Weisskopf et al. 2011). 
This source is found to be the only non-variable X-ray object without any optical/IR counterpart within its $\gamma$-ray error circle. 
Its association with the pulsar is reinforced by the fact that its X-ray position is consistent with the optimal $\gamma$-ray timing solution.

For firmly establishing the link between this X-ray source and \psr, a X-ray periodicity search and a detailed spectroscopy is necessary. 
Nevertheless, this is forbidden by the limited temporal resolution and the small photon statistics of the archival data. 
In view of this, we have investigated this Geminga-like pulsar with a deep {\it XMM-Newton} observation.
In this Letter, we report the discovery of the X-ray pulsation from \psr. This is the third member of this class with the significant 
pulsation detected from both X-ray and $\gamma-$ray regimes.

\section{Observations and data analysis}

We have observed \psr\ with {\it XMM-Newton} on 2012 April 11 for a total exposure of $\sim$ 133 ks (Obs.~ID: 0670590101; PI: Hui). 
The median satellite bore-sight pointing during this observation is RA=$20^{\rm h}21^{\rm m}30.56^{\rm s}$ Dec=$+40^{\circ}26^{'}46.8^{''}$ (J2000), which is the position
of 2XMM~J202131.0+402645 determined by Trepl et al. (2010). While the MOS1/2 CCDs were operated in full-window mode, PN CCD was operated in small-window mode with a temporal resolution of $\sim$5.7 ms, which enables the pulsation search for the first time. 
With the most updated instrumental calibration, we generated the event lists from the raw data obtained from all EPIC instruments with the tasks \emph{emproc} and \emph{epproc} of the XMM-Newton {\bf S}cience {\bf A}nalysis {\bf S}oftware (XMMSAS version 12.0.1). 
We selected only those events for which the pattern was between $0-12$ for MOS cameras and $0-4$ for the PN camera. 
We also noted that our data have been contaminated by the hard X-ray background flare. 
After removing all events which are potentially contaminated, the effective exposures are found to be 85~ks, 72~ks and 77~ks for MOS1, MOS2, PN respectively.

\subsection{Timing analysis}
For timing analysis, we utilized solely the PN data for the pulsation search. 
To determine the source position of 2XMM~J202131.0+402645, we have run the source detection with the XMMSAS task {\it edetect$\_$chain}. 
The source can be significantly detected with the signal-to-noise ratio of $\sim 72\sigma$ at (J2000) R.A.=$20^h21^m30^s.53$, decl.=$+40^{\circ}26'45''.5$ with the uncertainty $\sim 0''.4$.
We extracted the events within a circular region of a 20$''$ radius centered at this position, which corresponds to an encircle energy function of $\sim76\%$. 
Within an energy band of 0.15--12 keV, 3174 counts were available for the timing analysis. 
With the aid of the task {\it barycen}, the arrival times of all the selected events were barycentric corrected with the aforementioned X-ray position and the latest JPL DE405 earth ephemeris. 

Search around the rotational frequency of \psr\ with the method of epoch-folding results in a very significant detection of X-ray pulsation from 2XMM~J202131.0+402645. 
Using either Rayleigh-test (Mardia 1972; Gibson et al. 1982) or $H$-test (de Jager et al. 1989; de Jager \& B\"usching 2010),  a significant peak is found at the frequency 3.7689937(9)~Hz (=0.26532281(6)~s) with $Z_{1}^{2}=126$ and $H=133$ respectively.
This corresponds to the random probability of $< 10^{-14}$ (de Jager \& B\"usching 2010). 
The quoted uncertainty of our detected frequency is determined by following the method in Leahy et (1987).

Although the ephemeris of \psr\ has been determined several times by {\it Fermi} LAT observations at different epochs 
(Abdo et al. 2009a,2010a; Ackermann et al. 2011; Ray et al. 2011; Weisskopf 2011), none has covered the epoch of our {\it XMM-Newton} observation 
(i.e. $\sim$MJD~56028). In order to avoid the uncertainties introduced by extrapolating these timing models to our observation epoch, 
we derived a $\gamma-$ray ephemeris which is contemporaneous with this X-ray observation for comparing the temporal properties in both regimes. 

We obtained the {\it Fermi} LAT data in the energy range of 0.1-300 GeV within a circular region of interest (ROI) with a $0.8^{\circ}$ radius 
from the X-ray position of \psr. In order to achieve a reasonable signal-to-noise ratio and with the influence of the accumulated timing noise 
minimized, we adopted a data span of $\sim1$~year, MJD~55900-56280, which brackets the epoch the {\it XMM-Newton} observation. 
For data reduction, the {\itshape Fermi} Science Tools v9r23p1 package, 
available from the {\itshape Fermi} Science Support Center\footnote{http://fermi.gsfc.nasa.gov/ssc/data/analysis/software/}, was used.
We used Pass 7 data and selected events in the ``Source" class (i.e.~event class 2) only.
In addition, we excluded the events with zenith angles larger than 100$\degr$ to greatly reduce the contamination by Earth albedo $\gamma$-rays.
For determining pulse times of arrivals (TOAs), we firstly constructed a template based on the timing model reported by 
Ray et al. (2011) with 
the method of Gaussian kernal density estimation. By cross-correlating the template with unbinned geocentered data, which each photon 
assigned with a phase in accordance with the model reported by Ray et al. (2011), we measured the TOAs from 25 segments 
of adopted data span. 

Using TEMPO2, we fitted the TOAs to a timing model including $f$, $\dot{f}$, $\ddot{f}$ and $\dddot{f}$. All spin parameters 
are allowed to be free with the high order derivatives to account for most of the 
red noise. For the position, as the short data span does not allow a fit, we fixed it at the X-ray position determined by {\it Chandra} 
throughout the analysis (Weisskopf et al. 2011).
The results are summarized in Table~1. Since the high order frequency derivatives are dominated by the timing noise, we stress that 
their values do not reflect the intrinsic rotational properties of the pulsar. 
Using the parameters in Tab.~1, the extrapolated $f$ and $\dot{f}$ are consistent with the ephemeris at the specified epochs 
reported by earlier literatures within the statistical uncertainties.
According to this local ephemeris, the spin frequency of \psr\ at 
the start of the good time interval (MJD~56028.31153) of our {\it XMM-Newton} observation is 3.768993208(2)~Hz, 
which is consistent with the detected X-ray pulsation (i.e. $f=3.7689937(9)$~Hz).  
 
\begin{table*}
\begin{center}
\caption[]{Local ephemeris of \psr\ derived from LAT data which brackets the {\it XMM-Newton} observation on 
MJD~56028. The numbers in parentheses denote errors in the last digit.\\}
\begin{tabular}{ll}
\hline\hline
\multicolumn{2}{l}{Parameter} \\
\hline
Pulsar name\dotfill & \psr\ \\
Valid MJD range\dotfill & 55908---56274 \\
Right ascension, $\alpha$\dotfill & 20:21:30.733 \\
Declination, $\delta$\dotfill &  +40:26:46.04 \\
Pulse frequency, $f$ (s$^{-1}$)\dotfill & 3.768995206(2) \\
First derivative of pulse frequency, $\dot{f}$ (s$^{-2}$)\dotfill & $-8.166(3)\times10^{-13}$ \\
second derivative of pulse frequency, $\ddot{f}$ (s$^{-3}$)\dotfill & $-2(1)\times10^{-22}$ \\
Third derivative of pulse frequency, $\dddot{f}$ (s$^{-4}$)\dotfill & $1(2)\times10^{-29}$ \\
Epoch of frequency determination (MJD)\dotfill & 56000 \\
Solar system ephemeris model\dotfill & DE405 \\
Time system \dotfill & TDB \\
\hline
\end{tabular}
\end{center}
\end{table*}

\begin{figure}[!t]
\centerline{\psfig{figure=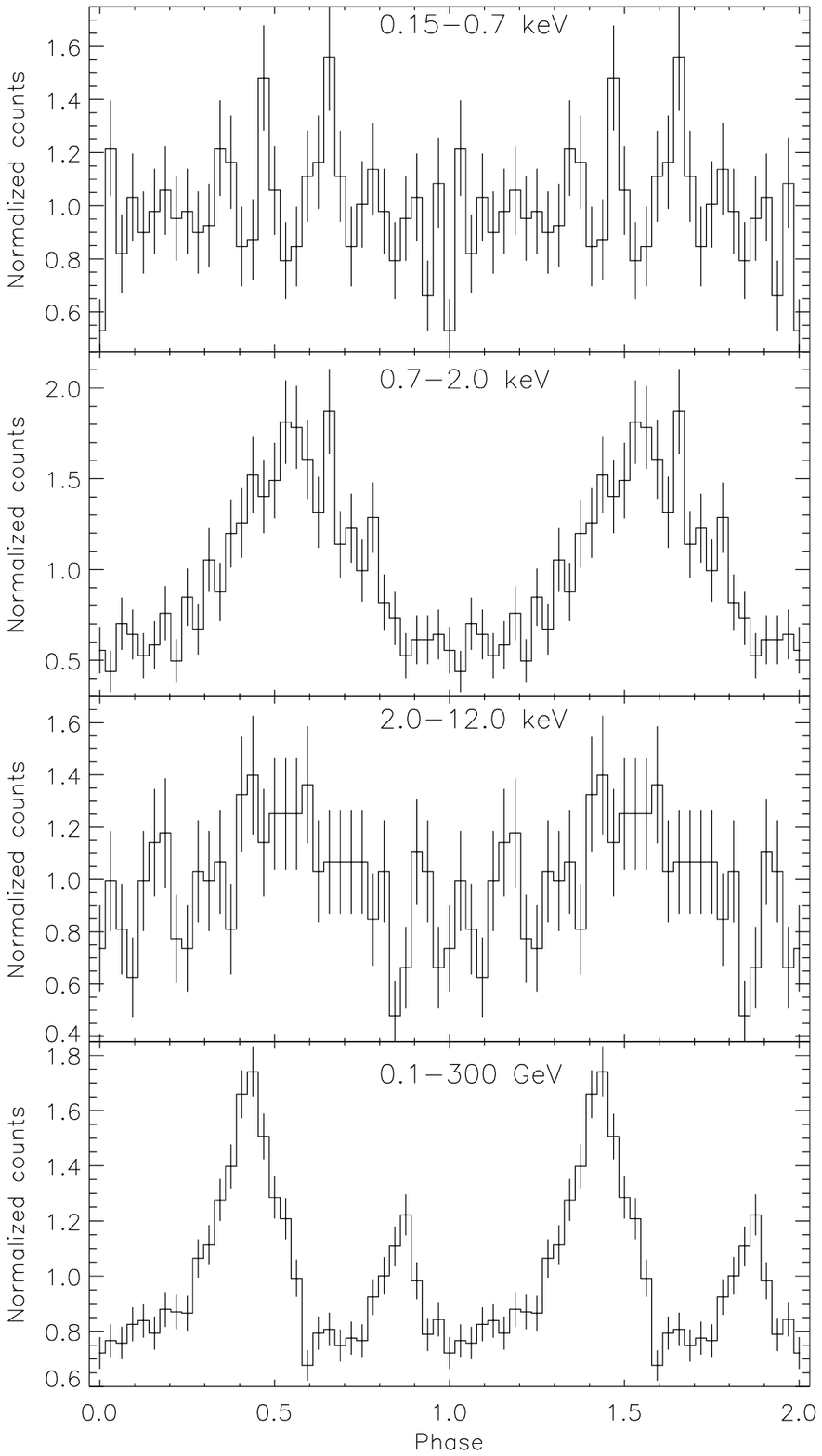, width=12cm,clip=,angle=0}}
\caption{Folded light curves of \psr\ in different energy ranges. Each panel shows the pulse profile of 32 bins in the energy band specified in the legend. 
All the light curves were folded with the timing parameters at the epoch/phase zero of MJD 56028 derived from Table~1.}
\label{PFXG}
\end{figure}
 
To compare the X-ray and $\gamma$-ray pulse profiles, the photons extracted from our {\it XMM-Newton} observation and 
those obtained from {\it Fermi} within MJD 55910 to 56110 were all folded in accordance with our derived $\gamma-$ray ephemeris (i.e. Tab.~1) 
with the epoch zero at MJD~56028.
For further examining the properties of the X-ray pulsation, we performed the energy-resolved timing analysis by dividing 
the X-ray data into three consecutive bands: 0.15--0.7~keV (soft), 0.7--2~keV (medium) and 2--12~keV (hard). 
The phase-aligned pulse profiles at different energy ranges are shown in Fig.~\ref{PFXG}. 
Apparently, the X-ray pulsation is most significant in the medium band ($Z^{2}_{1}=166$). 
In the hard band, there is a very marginal pulse detection of
$Z^{2}_{1}=17.8$. However, this corresponds to a much higher random probability of $\sim8\times10^{-4}$.
In the soft band, there is no conclusive evidence for the pulsation.

\begin{figure*}[t]
\begin{center}
{\psfig{figure=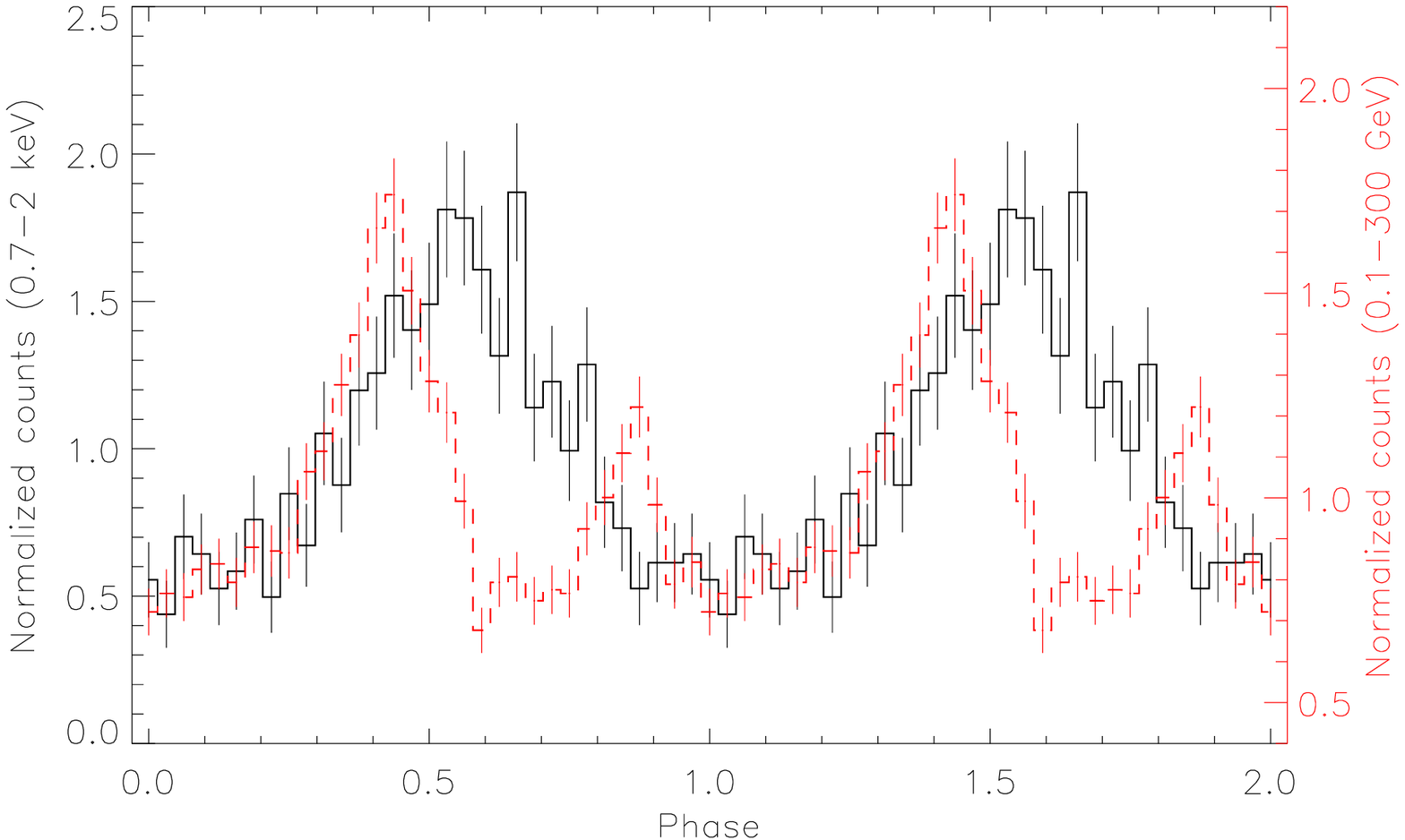, width=8.1cm,clip=,angle=0}}{\psfig{figure=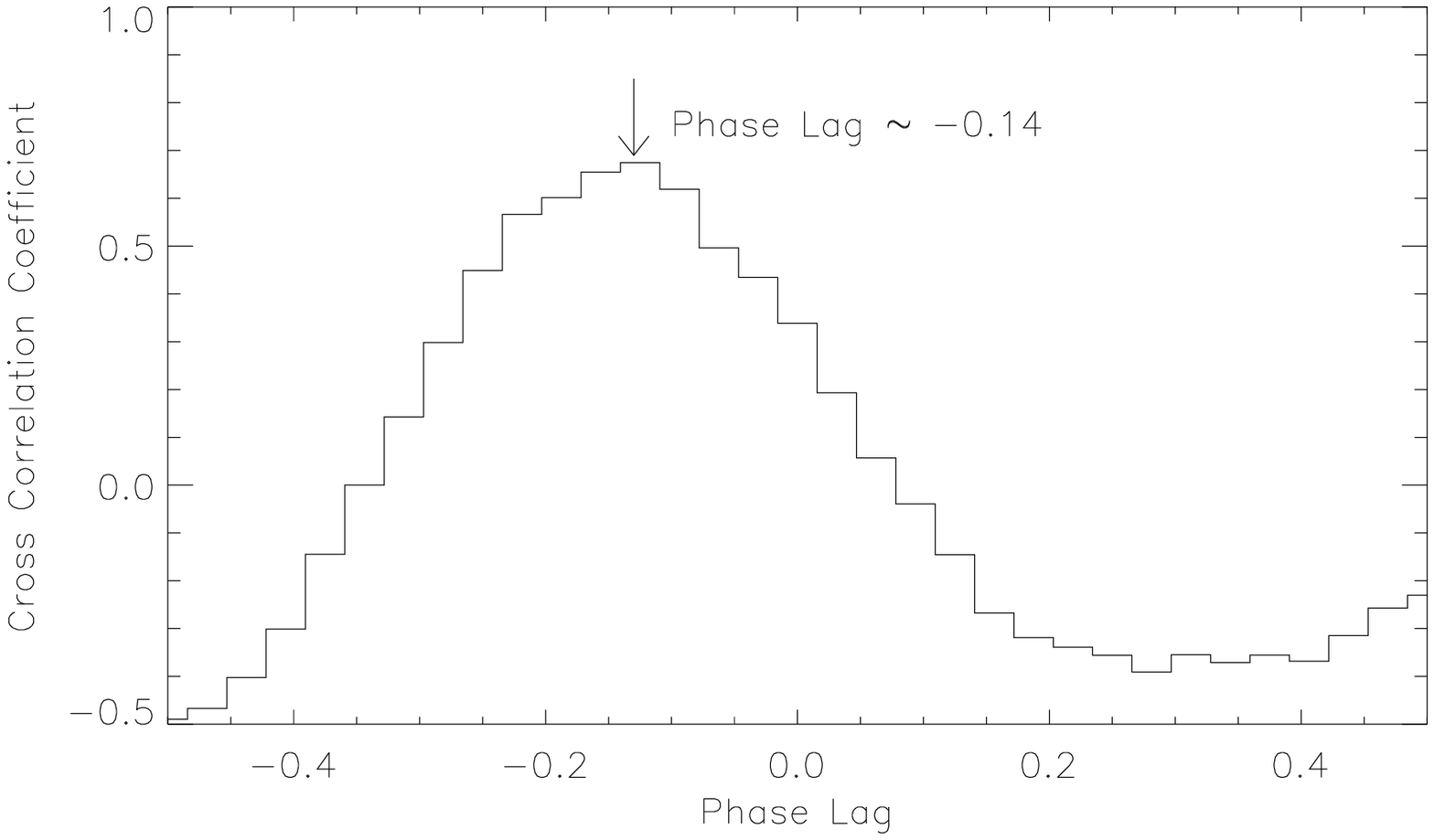, width=8.1cm,clip=,angle=0}}
\end{center}
\caption{Correlation of the X-ray and $\gamma$-ray pulsations. {\it Left panel:} The light curves of \psr\ in 
X-ray and $\gamma$-ray bands both folded at the epoch zero of MJD 56028 with 32 bins using the same ephemeris reported in this letter. 
The pulse profiles in 0.7--2~keV and 0.1--300~GeV are illustrated as solid line and dashed line respectively.  
{\it Right panel:} Cross-correlation of the X-ray and $\gamma$-ray pulse profiles is shown. The coefficient has 
a maximum value at a phase lag of -0.14.} 
\label{Correlation}
\end{figure*}

There is a recognizable phase offset between the $\gamma$-ray and the X-ray profiles (see the left panel in Figure~\ref{Correlation}). 
To quantify the phase offset, we computed the cross-correlation between these two phase-aligned light curves.
The result is shown in the right panel of Figure~\ref{Correlation}.
The cross-correlation coefficient attains the maximum value at a phase-lag of $-0.138$. 
For estimating the uncertainty of cross-correlation, we used Monte Carlo simulation to obtain the distribution of phase-lag. 
The simulated X-ray light curves were generated by the original one plus a random series of variates sampled from a 
Gaussian distribution with a mean of 0 and a standard derivation of $1\sigma$ corresponds to the errors of observed 
folded light curve. With 10000 trials, the standard deviation is found to be 0.024 which suggests the aforementioned phase-lag is detected 
at a confidence level of $>5\sigma$.   
   
\subsection{Spectral analysis}
Apart from detecting the X-ray pulsation, our deep {\it XMM-Newton} observation is also able to place a 
tight constraint on the spectral properties of \psr. 
For spectral analysis, we utilized the data obtained from all three EPIC cameras. 
With the aid of the XMMSAS task {\it epatplot}, all the EPIC data are found to be not affected by CCD pileup. 
We extracted the spectrum of \psr\ from circles with a radius of $20''$ centered at its nominal X-ray position 
(see \S2.1) in MOS1, MOS2 and PN cameras respectively. 
The background spectra were extracted from the nearby regions in the corresponding CCDs, 
which are source-free and with sufficient size to enable a less biased sampling. 
The response files were generated by the XMMSAS task {\it rmfgen} and {\it arfgen}. 
After background subtraction, we have 1159~cts collected by EPIC in the energy range of 0.5-10 keV. The photon statistic is $\sim4$ times 
higher than that used in the spectral analysis reported by (Weisskopf et al. 2011). 
We grouped each spectrum dynamically so as to achieve the same signal-to-noise ratio in each dataset. 
In order to tightly constrain the spectral parameters, we fitted the data obtained from three cameras simultaneously with XSPEC 12.6.0. 
For a conservative estimation of uncertainties, the quoted errors of the spectral parameters 
are $1\sigma$ for 2 parameter of interest (i.e. $\Delta\chi^{2}=2.30$ above the minimum) for 
the single component models and $1\sigma$ for 4 parameter of interest (i.e. $\Delta\chi^{2}=4.72$ above the minimum) for the 
multi-component models.

With the tested single component models, we found that the pulsar spectrum cannot be appropriately described by these simple scenarios. 
For the absorbed blackbody, it results in a relatively poor goodness-of-fit ($\chi^{2}=65.33$ for 46 d.o.f.). 
On the other hand, although the absorbed power-law model results in an acceptable goodness-of-fit ($\chi^{2}=52.78$ for 46 d.o.f.), 
it yields a very large photon index ($\Gamma=4.4^{+0.6}_{-0.5}$) which is far steeper than that expected for a pulsar 
(cf. Cheng \& Zhang 1999). 
In both of these single component fits, systematic deviations have been noted in the fitting residuals for the energies larger than $\sim3$~keV. 
All these suggest the X-ray emission of \psr\ might consist of more than one spectral component. 

\begin{figure}[t]
\centerline{\psfig{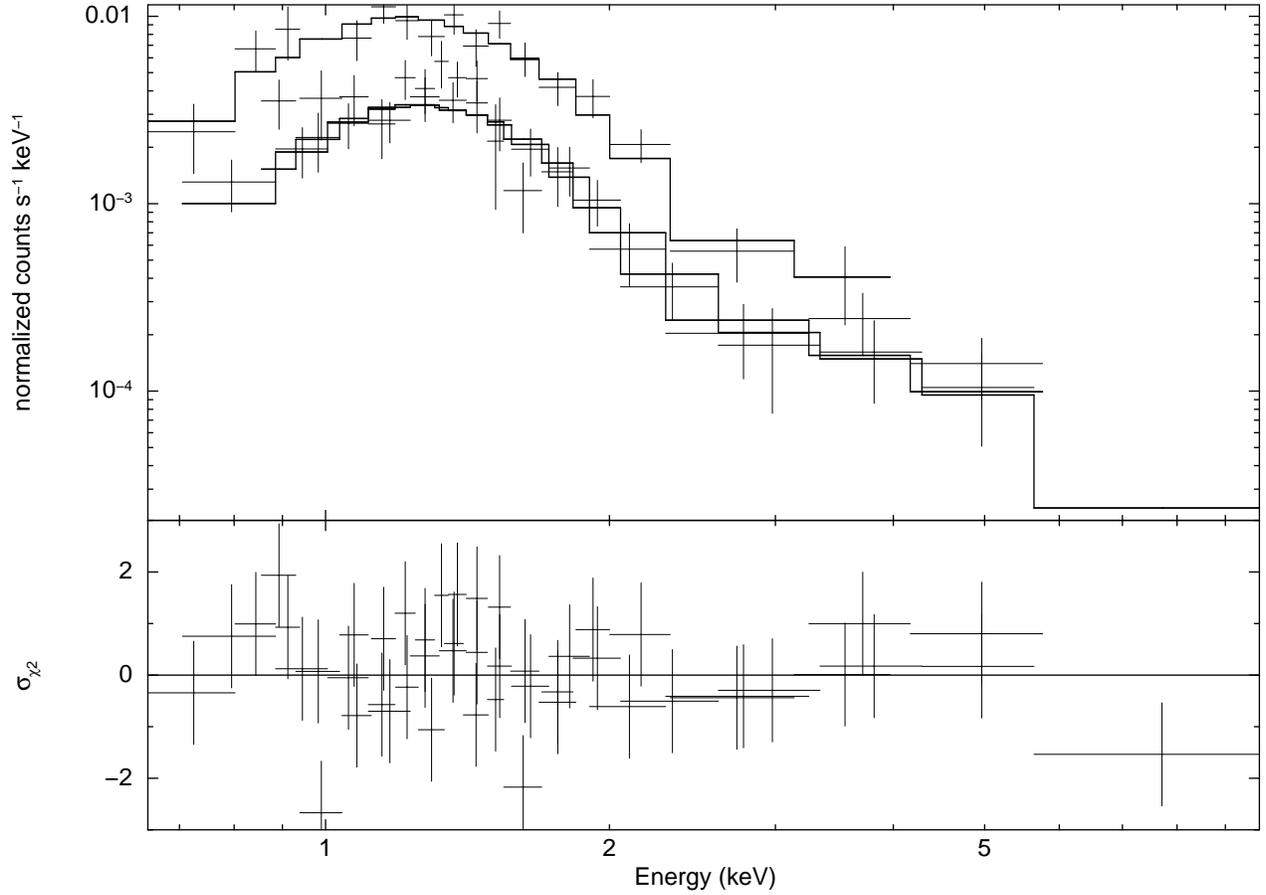}}
\caption[phase-averaged energy spectrum]{The phase-averaged energy spectrum of \psr\ in 0.5--10~keV. The X-ray emission from the position of \psr\ as observed with the PN (upper spectrum) and MOS1/2 detectors (lower spectra) are simultaneously fitted to an absorbed blackbody plus power-law model ({\it upper panel}) and contributions to the $\chi^{2}$ fit statistic ({\it lower panel}).}
\label{psr_spec}
\end{figure}

We proceeded to examine the spectrum with multi-component models. 
We found that the blackbody plus power-law model (BB+PL) can describe the data reasonably well ($\chi^{2}=41.55$ for 44 d.o.f.). 
It yields a column density of $n_{\rm H}=(7.0^{+4.1}_{-2.7})\times10^{21}$~cm$^{-2}$, a temperature of $kT=0.24\pm0.06$~keV, a 
blackbody emitting region with a radius of $R=251^{+537}_{-132}d_{1.5}$~m, a photon index of $\Gamma=1.2^{+1.7}_{-1.2}$ and a normalization for the power-law component at 1 keV of $(2.8^{+14.1}_{-2.6})\times10^{-6}$~photons~keV$^{-1}$~cm$^{-2}$~s$^{-1}$, 
where $d_{1.5}$ represents the distance to \psr\ in unit of 1.5~kpc. 
The best-fit BB+PL model and the residuals for the X-ray spectrum of \psr\ are shown in Figure~\ref{psr_spec}. 
On the other hand, a pure thermal scenario consists of two blackbody components (BB+BB) also results in a comparable goodness-of-fit ($\chi^{2}=40.19$ for 44 d.o.f.). 
It yields a set of best-fit parameters of $n_{\rm H}=(6.7^{+3.6}_{-2.5})\times10^{21}$~cm$^{-2}$, $kT_{1}=1.4^{+1.8}_{-0.6}$~keV, 
$R_{1}=3.6^{+6.4}_{-2.5}d_{1.5}$~m, $kT_{2}=0.25\pm0.05$~keV and $R_{2}=223^{+320}_{-106}d_{1.5}$~m. 
Both of these composite models imply an unabsorbed flux at the level of $\sim10^{-13}$~ergs~cm$^{-2}$~s$^{-1}$. 

\section{Discussion}
In this Letter, we report our detection of X-ray pulsation from 2XMM~J202131.0+402645. 
The detected spin frequency is consistent with the $\gamma$-ray pulsation of \psr\ at the same epoch. 
This provides a clear evidence that the observed X-rays are indeed from \psr. 

The X-ray pulse profile in 0.7-2~keV resembles the modulation
resulted from a rotating neutron star with a hot spot on the stellar surface sweeps across our line-of-sight 
(e.g. Hui \& Cheng 2004; Pechenick et al. 1983).
This is consistent with the scenario inferred from the phase-averaged spectral analysis which favors 
a model of blackbody plus power-law or a composite blackbody model. 
In both of these best-fit models, a blackbody component with a temperature of $kT\sim0.25$~keV is required. 
At a distance of 1.5~kpc, the best-fit blackbody radius is found to be $\sim0.2-0.3$~km. 
This is comparable with the canonical size of a polar cap, $R\sqrt{\frac{R\omega}{c}} \sim 280$~m, where $R\sim 10$~km 
represents the typical radius of a neutron star and $\omega$ is the angular frequency of \psr. 
According to the outer gap model (Takata, Cheng, \& Taam 2012; Cheng \& Zhang 1999), the temperature of the polar cap heated by the return current 
injected by the gap is $kT\sim0.3$~keV which is also consistent with the observed value.  
Furthermore, this spectral component contributes $>80\%$ in the observed flux in both BB+PL and BB+BB fits in 0.7-2~keV. 
All these results point to the thermal nature of the observed X-ray pulsation. 
 
In the BB+PL fit, the non-thermal component with a photon index of $\Gamma\sim1.2$ is typical for a canonical pulsar.  
One possible origin of these non-thermal X-rays is the synchrotron radiation of the relativistic 
$e^{-}/e^{+}$ from the outer magnetospheric gap (Cheng \& Zhang 1999; Takata et al. 2006,2008), which should give rise to a pulsed signal. 
Despite the best-fit power-law component contributes $>90\%$ in the observed flux at the energies $>2$~keV, 
the evidence for the pulsation in the hard band is marginal.
This suggested the PL component might be non-pulsed in nature, which can possibly be originated from a pulsar wind nebula (PWN). 
Indeed, a putative bow-shock associated with \psr\ has been marginally resolved by a {\it Chandra} observation 
(Weisskopf et al. 2011). It might contribute an steady unpulsed non-thermal X-ray emission 
as a DC level which can be found in all rotational phases. 

On the other hand, the BB+BB model is also able to fit the observed data with a comparable goodness-of-fit. 
While the low temperature component is consistent with that inferred in the BB+PL fit, it also requires 
a hotter component with a much smaller emitting area with a radius of few meters. 
This might be a phenomenological two-steps adaptation for a wider temperature distribution of the hot polar cap 
with the hotter component describing the peak of the distribution. 
The marginal detection of X-ray pulsation in the hard band might be originated from the modulation of this component.
However, the inferred temperature, $\sim2\times10^{7}$~K, is higher than that expected from the heating 
by the back flow current from the outer gap (Cheng \& Zhang 1999) unless the surface multipolar magnetic field 
is strong enough to reduce the pure dipolar cap area. 

Comparing the phase-aligned X-ray and $\gamma$-ray light curves (Fig.~\ref{PFXG}), their pulse morphologies are clearly different where the $\gamma$-ray profile is narrower and contains at least two peaks.
In cross-correlating these two profiles, the cross-correlation coefficient is found to attain the maximum value 
of $\sim 0.67$ at a phase lag of $\sim-0.14$ as shown in the right panel of Fig.~\ref{Correlation}. 
This clearly indicates that the observed X-rays and $\gamma$-rays are originated from different sites. By comparing these observed results with the model predictions,  the pulsar emission geometry (e.g. magnetic inclination angle, viewing angle) can be constrained. 
In the previous study, Trepl et al. (2010) has modelled the $\gamma$-ray pulse profile of \psr\ in the context of the outer gap model. 
Nevertheless, due to the lack of pulses detected in radio and any other wavelength at that time, there was an uncertainty in determining which peak is leading (see Fig.~13 \& 14 in Trepl et al. 2010). 
With our detection of X-ray pulsation, such ambiguity can now be resolved and thus the theoretical investigation of this 
pulsar can be re-visited.

\acknowledgments

This research has made use of the data obtained through the {\it Fermi} Science Support Center Online Service, provided by the NASA/Goddard Space Flight Center. 
The work is partially supported by the National Science Council through grants NSC 101-2112-M-039-001-MY3. 
C.Y.H. and K.A.S. are supported by the National Research Foundation of Korea through grant 2011-0023383.
C.P.H. and Y.C. are supported by the National Science Council of Taiwan through the grant NSC 101-2112-M-008-010.
R.H.H.H. acknowledges support from the National Science Council through grants NSC 100-2628-M- 007-002-MY3 and NSC 100-2923-M-007-001-MY3. 
J.T. and K.S.C. are supported by a GRF grant of HK Government under HKU700911P.

\end{document}